\documentclass[10pt]{article}
\usepackage[left=2.5cm, right=2.5cm, top=2.5cm, bottom=2.5cm]{geometry}

\usepackage{graphicx}
\usepackage{authblk}
\usepackage{mathtools}
\usepackage{amssymb,amsmath,amsfonts,amsthm}
\usepackage{enumitem}
\usepackage{setspace}
\usepackage{hyperref}
\hypersetup{colorlinks=true, linkcolor=blue, citecolor=blue, urlcolor=blue}

\usepackage{changepage}
\usepackage{fancyhdr}

\usepackage{multirow}
\usepackage{array}
\usepackage{booktabs}
\usepackage{caption}
\usepackage{subfigure}
\usepackage{float}
\usepackage{wrapfig}
\usepackage{longtable}
\usepackage{tabularx}
\usepackage{colortbl}

\usepackage{algorithm}
\usepackage{algpseudocode}

\usepackage{amsbsy}
\usepackage{mathrsfs}
\usepackage{tipa} 
\usepackage{latexsym}

\usepackage{listings}
\usepackage[normalem]{ulem}

\usepackage{tikz}
\usetikzlibrary{calc}

\usepackage[symbol]{footmisc}
\usepackage[sort&compress,numbers,square]{natbib}
\usepackage{doi} 

\title{Filling of incomplete sinograms from sparse PET detector configurations using a residual U-Net}

\author[1,2]{Klara Leffler\footnote{Corresponding author: kl@math.ku.dk}}
\author[3]{Luigi Tommaso Luppino\footnote{Luigi Tommaso Luppino is now at Norsk Regnesentral, Oslo, Norway.}}
\author[3,4,5]{Samuel Kuttner}
\author[6]{Karin Söderkvist}
\author[7]{Jan Axelsson}

\affil[1]{Department of Mathematics and Mathematical Statistics, Umeå University, Sweden}
\affil[2]{Department of Mathematical Sciences, University of Copenhagen, Denmark}
\affil[3]{UiT Machine Learning Group, Department of Physics and Technology, UiT The Arctic University of Norway, Norway}
\affil[4]{The PET Imaging Center, University Hospital of North Norway, Norway}
\affil[5]{Nuclear Medicine and Radiation Biology Research Group, Department of Clinical Medicine, UiT The Arctic University of Norway, Norway}
\affil[6]{Department of Diagnostics and Intervention, Oncology, Umeå University, Sweden}
\affil[7]{Department of Diagnostics and Intervention, Radiation Physics, Umeå University, Sweden}

\date{\normalsize \today}

\begin{document}

\maketitle

\vspace*{-0.5cm}
\begin{abstract}
\noindent {\bf Background:} 
Long axial field-of-view PET scanners are becoming increasingly available worldwide for clinical and research nuclear medicine examinations, providing an increased field-of-view and sensitivity compared to traditional PET scanners. However, a significant cost is associated with manufacturing the densely packed photodetectors required for the extended-coverage systems. Despite improved performance allowing ultralow dose or ultrafast scans, the financial barrier remains, limiting clinical utilisation. \\ 
{\bf Purpose:} 
To mitigate the cost limitations, alternative sparse system configurations with strategically placed inter-detector gaps have been proposed, allowing an extended field-of-view PET design with detector costs similar to a standard PET system, albeit at the expense of image quality.\\
{\bf Methods:} 
To address the challenges posed by sparse detector configurations, particularly the heavy undersampling of PET measurements, we propose a deep sinogram restoration network to fill in the missing sinogram data. The network, a modified Residual U-Net, is trained end-to-end using standard clinical PET scans from a GE Signa PET/MR. The training involves simulating the removal of 50\% of the detectors in chessboard patterns of varying sizes, leading to incomplete sinograms with significant count losses (thus retaining only 25\% of all lines of response).  \\
{\bf Results:}
The model successfully recovers missing counts in incomplete sinograms, with a mean absolute error consistently below two events per pixel for typical injected radioactivity, outperforming 2D interpolation of incomplete sinograms based on mean absolute error and structural similarity in both sinogram and reconstructed image domain. 
Notably, the predicted sinograms exhibit a smoothing effect, leading to reconstructed images lacking sharpness in finer details. 
Despite these limitations, the model demonstrates a substantial capacity for compensating for the undersampling caused by the sparse detector configuration. \\
{\bf Conclusions:} 
This proof-of-concept study suggests that sparse detector configurations, combined with deep learning techniques, offer a viable alternative to conventional PET scanner designs. This approach supports the development of cost-effective, total body PET scanners, allowing a significant step forward in medical imaging technology.\\
\end{abstract}

\medskip
\noindent Keywords: 
positron emission tomography (PET), 
sparse PET, 
sinogram restoration, 
deep learning -- artificial intelligence

\section{Introduction}
Positron Emission Tomography (PET) imaging has become a cornerstone in medical diagnostics. Technology-driven developments such as iterative image reconstruction and increased computational power have improved performance and paved the way for constructing long axial field-of-view (FOV) "total" body PET systems. Current commercial models like the United Imaging uExplorer\textsuperscript{TM} total-body PET scanner and the Siemens Biograph Vision Quadra\textsuperscript{TM} nearly total-body PET scanner, offering axial FOVs of up to 194 cm and 106 cm, respectively, are examples of this latest technological leap~\citep{uExplorer, Vision}. The updated PennPET Explorer further exemplifies this trend towards extended FOV~\citep{PennPET, PennPET2, PennPETex}. These advancements promise benefits like lower radioactive doses, shorter scan times, and reduced image noise. However, they come at a significant cost, primarily due to the increased number of detectors required~\citep{Vandenberghe2020, Abgral2021, Daube-Witherspoon2021, Karakatsanis2022}. Therefore, the traditional cylindrical configuration of PET scanners, characterised by densely packed photodetectors, presents substantial financial limitations in the transition to extended FOV PET imaging. 

A possible, cost-effective solution is to introduce gaps between detector elements, either in the axial or transaxial direction or a combination thereof. Such a system geometry potentially extends detector coverage without requiring additional crystal material, thereby providing an extended FOV scanner with reduced material requirements compared to conventional models~\citep{Daube-Witherspoon2021, Knopp2019, ZeinKarakatsanis2020, Karp2022, Vandenberghe2023}. However, the sparse detector configurations invariably affect image quality, leading to overall sensitivity loss, increased background variability, and reduced spatial resolution and contrast recovery~\citep{Daube-Witherspoon2021, Yamaya2008, Yamaya2009, AbiAklVandenberghe2019,  ZeinKarakatsanis2021}. Simulation-based comparisons have confirmed that such configurations can reduce system cost but substantially lower sensitivity, especially in designs with aggressive sparsity levels~\citep{dadgar2023simulation, gao2023cost_effective}. Nonetheless, cost-effectiveness analyses suggest that extended FOV PET systems may still be financially justifiable in select clinical applications, particularly when paired with algorithmic compensation strategies~\citep{alberts2025costeffect}.
~
Various compensatory methods have been explored to address challenges related to detector gaps, including compressed sensing and inpainting~\citep{Ahn2012, Valiollahzadeh2015a, Shojaeilangari2018}, sinogram interpolation~\citep{Karakatsanis2018}, optimisation-based reconstruction algorithms~\citep{Zhang2019}, normalisation to account for missing lines of response (LORs)~\citep{Knopp2019, ZeinKarakatsanis2020, gao2023cost_effective}, and continuous bed motion~\citep{Karakatsanis2022, ZeinKarakatsanis2021, Karakatsanis2019}. However, these solutions have often been applied to small-scale data losses and struggle to handle the massive count losses that arise from removing large portions of detectors elements from the PET configurations. The normalisation approach, although offering a large-scale solution, has previously only been applied to either axial or transaxial gaps of small sizes and has yet to be successfully implemented on the complex missing data patterns resulting from sparsity patterns that combine axial and transaxial gaps. 
Furthermore, while continuous bed motion helps average out undersampling artifacts, it does not address the root problem of missing detector elements and remains available only on select scanners, limiting its applicability in sparse detector configurations.

In recent years, deep learning has emerged as a potent tool in medical imaging, particularly for data completion tasks like estimating high-count PET data from low-count measurements~\citep{Zaidi2020, Zaidi2021, WangDaldrup-Link2023}. A recent review highlights the expanding role of deep learning across the PET reconstruction pipeline~\citep{hashimoto2024review}. The detector gap problem has been studied as a learning problem in smaller scales such as a few uniformly spaced axial gaps~\citep{Sheikhzadeh2019, Amirrashedi2021, Whiteley2019, LuFeng2021}, or a C-shaped ring design~\citep{Huang2019}. However, these studies did not address the extensive and complex patterns of missing data that arise in more aggressive sparse PET configurations, such as those explored here. Moreover, most deep learning applications in PET reconstruction have focused on the image domain; direct sinogram restoration from sparsely sampled data remains largely unexplored.

Therefore, this proof-of-concept study proposes a novel deep learning-based approach to fill incomplete sinograms resulting from inter-detector gaps when removing 50\% of detectors in a chessboard pattern. Our method, leveraging a deep sinogram restoration network, presents a practical and efficient solution to address the significant count losses and image quality degradation inherent in sparse PET system configurations. This approach enhances the feasibility of cost-effective extended FOV PET scanners and represents a novel technical advancement in the application of artificial intelligence in nuclear medicine.

\section{Methods}\label{sec2}
\subsection{Data description and processing}
\begin{figure}[ht]
    \centering
    \captionsetup{width=.91\linewidth}
    \includegraphics[width=.5\linewidth]{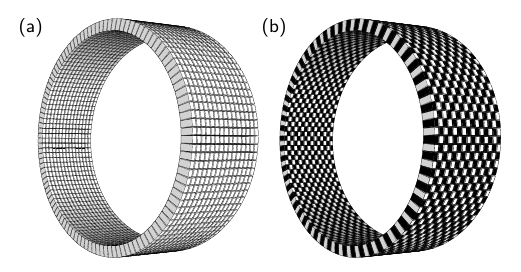}
    \caption{Example comparison of a standard compact PET configuration 
        \textbf{(a)} and a sparse PET configuration using a chessboard detector pattern 
        \textbf{(b)}. For the sake of resolution, the shown mockup scanners comprise of 15 rings with 128 detector elements each.
        \label{fig:sparsePET}
        }  
\end{figure}

The PET dataset originates from a clinical trial, MORRIS (NCT02379039), approved by the Regional Ethical Review Board, Ume{\aa} (dnr 2015/117-31). 
To evaluate the proposed method, 8 different pelvis scans acquired without time-of-flight on a GE Signa PET/magnetic resonance (MR) imaging system were used. 
The original dataset, therefore, contained information from all detector positions. 
To simulate a sparse PET configuration, a subsampling scheme cancelling half of the detectors in a chessboard pattern was adopted (Figure~\ref{fig:sparsePET}). 
The chessboard pattern was created by cancelling single crystals units, i.e. $1\times1$ crystals in each chessboard square, resulting in gaps of size $4\times5.3$mm. 
Detector removal was performed by discarding coincidences from the original list mode data file and binning the new list mode file into 1981 2D sinograms per scan (the scanner's standard sinogram format). 
The resulting count loss corresponds to deleting approximately 75\% of the lines of response. In the sparse configuration, we evaluated cancelling out crystal units corresponding to both black and complimentary white positions as shown in Figure~\ref{fig:sparsePET}. 
The modified dataset contained $8\times2\times1981=31696$ incomplete 2D sinogram slices, counting 8 individual scans, chessboard and complimentary chessboard patterns, and all 1981 direct and oblique planes.

\begin{figure}[ht]
    \centering
    \captionsetup{width=.91\linewidth}
    \includegraphics[width=.55\linewidth]{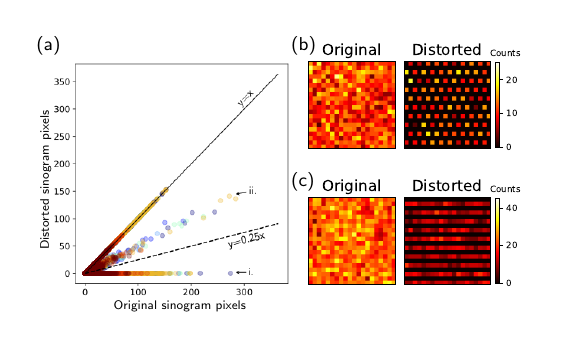}    \caption{Visualisation of the sinogram distortion due to the sparse PET chessboard configuration. 
        \textbf{(a)} Pixel-wise correlation between the original and distorted sinograms; figure shows random sample of $10^5$ pixels from the different scans (different colours for different scans). 
        The distorted pixels follow two separate distributions (marked by arrows) -- they are either zero (i) or of lower intensity than the original (ii). 
        The ideal fit is illustrated by a solid line, and the overall total fit to the original pixels is illustrated by a dashed line.
        \textbf{(b)} shows the sinogram distortion pattern with zeroed out pixels in direct planes and cross planes with ring difference $>1$. 
        \textbf{(c)} shows the sinogram distortion pattern with low-intensity pixels in summed cross planes with ring difference 1.
        \label{fig:Sino_scatterplots_in}
        }
\end{figure}

One can determine which lines of response are associated with removed detectors and, therefore, create a binary sinogram mask indicating affected sinogram pixels. 
Highlighted pixels in the mask must be restored, whereas the others can be left as they are. 
The effects of the detector cancellation caused sinogram pixels to either be zeroed out or have lower intensity specific 2D sinograms (Figure~\ref{fig:Sino_scatterplots_in}(a)). 
The main pattern for missing counts corresponded to a grid of diagonal zero-valued lines (Figure~\ref{fig:Sino_scatterplots_in}(b)). 
The second pattern containing low-intensity values and zero-valued lines appeared in added cross planes for ring differences equal to one (Figure~\ref{fig:Sino_scatterplots_in}(c)).

\subsection{Sinogram restoration network}
\begin{figure*}[h!]
    \centering
    \captionsetup{width=.91\linewidth}
    \includegraphics[width=1\linewidth]{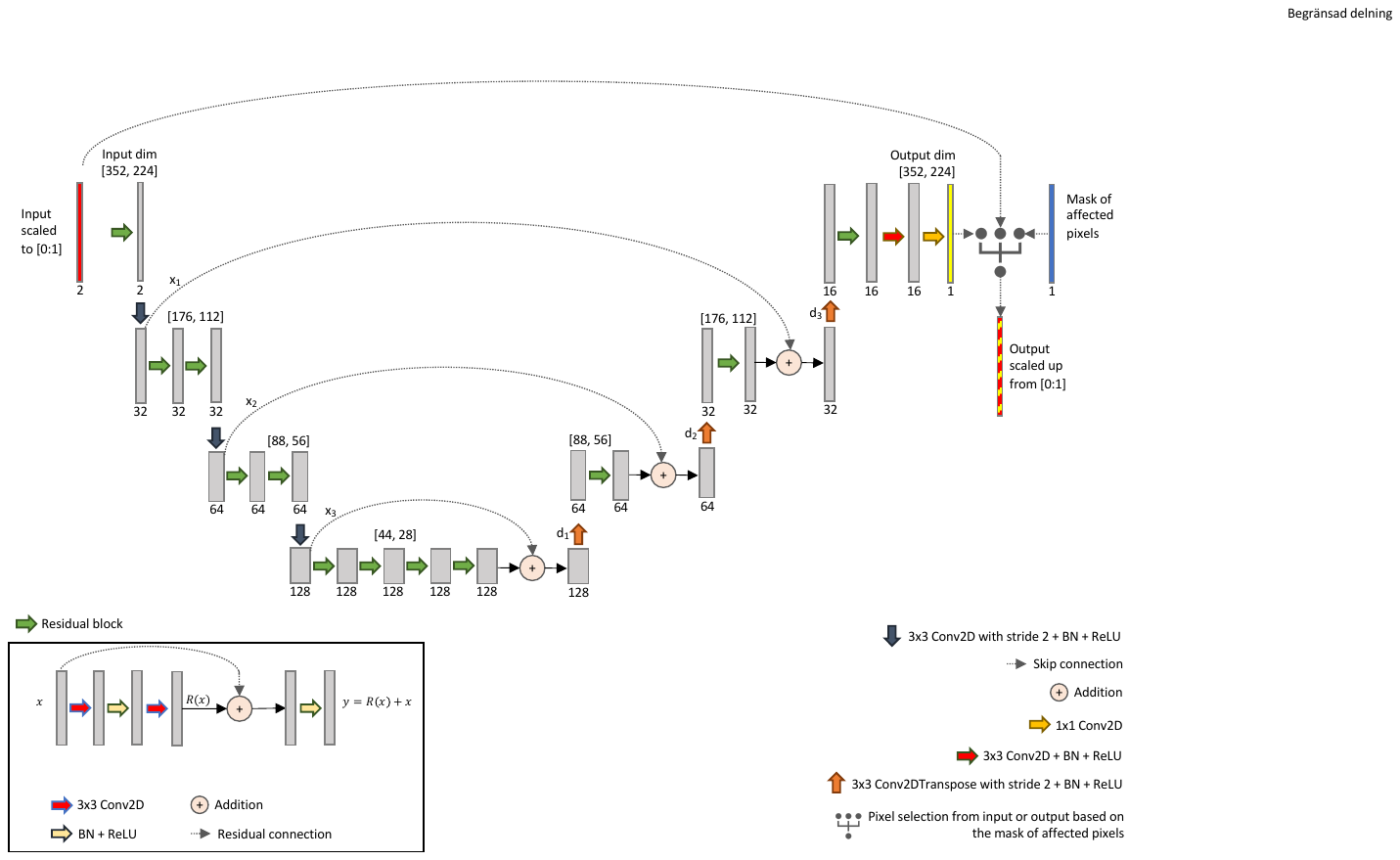}
    \caption{Architecture of the implemented sinogram restoration network. 
        Each green arrow represents a residual block containing convolution, batch normalisation and rectified linear unit (ReLU) activation, as well as a residual connection. 
        The dotted arrows represent skip connections, and the circled plus signs represent addition of the features from the down-sampling layers with the up-sampling layers. 
        The network down-samples via convolution with stride 2, reducing the spatial dimensions by half, and up-samples using transpose convolution, doubling the spatial dimensions. 
        Two final convolutions layers are used to generate the output. The sinograms are normalised to range $[0:1]$ before input and denormalised by the same factor after output. 
        Finally, the pixels not affected by the removed detectors are copied from the input sinograms and reinstated in the output.
        \label{fig:ResUnet}
        }
\end{figure*}

The proposed deep network was inspired by a segmentation task~\citep{Myronenko2019} and consists of a residual U-Net~\citep{zhang2018road} combining the strengths of residual learning~\citep{he2016deep} and the well-known U-Net architecture~\citep{U-net}. 
The network consists of an asymmetric encoder-decoder structure (Figure~\ref{fig:ResUnet}). 
The encoder layers feature an increasing number of residual blocks and one strided convolution, which halves the spatial dimensions. 
The layers of the decoder are composed of strided transposed convolutions which double the spatial dimensions, followed by one residual block. 
Skip connections connect the down-sampling and up-sampling segments, allowing the network to combine positional and contextual information to predict the missing sinogram information and aid gradient back-propagation during training. 
Each sinogram was normalised to the range $[0:1]$ before input and then denormalised by the same factor after output to control the unboundedness of the sinogram values. 
This was shown to be beneficial during training, achieving a more steady and fast convergence~\citep{lecun2002efficient}. 
Finally, the pixels not affected by the removed detectors were copied from the input sinograms and reinstated in the output.

\subsection{Network training}
The proposed sinogram restoration network was trained on six patients ($6\times2\times1981=23772$ sinograms) at a time, holding out one patient for validation ($1\times2\times1981=3692$ sinograms) and one for testing ($1\times2\times1981=3692$ sinograms). 
Eight-fold cross-validation allowed assessment of model performance on all patients. 
A combined pixel-wise structural similarity index measure (SSIM) and $\ell_1$ based loss function $\mathcal{L}$ was used for the preservation of pixel accuracy as well as visual coherence. 
Specifically, we defined our loss as
\begin{align*}
\mathcal{L}(\hat{y},y)= (1-SSIM) + MAE
\end{align*}
where
\begin{align}
SSIM &= \frac{(2\mu_{\hat{y}}\mu_y + C_1)(2\sigma_{\hat{y}y} + C_2)}{(\mu_{\hat{y}}^2 + \mu_y^2 + C_1)(\sigma_{\hat{y}}^2 + \sigma_y^2 + C_2)}
 \label{eq:SSIM} \\
MAE &= \frac{1}{n}\sum_{i\in m} \vert \hat{y}_i-y_i \vert. \label{eq:MAE}
\end{align}
Here, $\hat{y}$ and $y$ are the predicted and original sinograms, respectively. The means $\mu$, variances $\sigma^2$ and covariance $\sigma_{\hat{y}y}$ are computed locally over corresponding pixels. The constants $C_1$ and $C_2$ depend on the dynamic range $L$ of the sinograms or images such that $C_1=(0.01L)^2$ and $C_2=(0.03L)^2$. The MAE is computed over a masked region $m$ consisting of $n$ activated pixels.

The loss function was minimised using the Adam optimiser~\citep{reddi2018on} for 200 epochs with a learning rate of $10^{-3}$, which experienced a stair-case exponential decay rate of 0.96. 
During each training epoch, batches of 64 sinograms were used until the whole training set was exhausted. The training was evaluated using the validation set at the end of each epoch. 
If the validation loss did not improve for 20 consecutive epochs, an early stopping criterion was triggered, effectively stopping the training.

The model was developed using the Keras modules of TensorFlow and trained on a high-performance computer equipped with 48 CPU cores and 2 NVIDIA A100 GPUs. The data was converted into the TFRecords format, and the loading pipeline was optimized, so each epoch took about 36 seconds. The minimum, maximum, and average training time (number of epochs) were 73 minutes, 122 minutes, and 98 minutes (109, 194, 149 epochs), respectively. Once the model was trained, the runtime to infer the restored sinograms was approximately of 20 seconds per patient. The same operation took approximately 2 and a half minutes on an Apple MacBook Pro M1 (2021), with a much more modest hardware configuration which is comparable to the ones found within typical clinical settings.

\subsection{Quantification}
The sinogram restoration network performance was evaluated in both sinogram and image domains. 
For comparison, interpolated 2D sinograms were also created using the Clough-Tocher method~\citep{Alfeld1984, Farin1986}. 
Both the network restored (all $1981\times2$ slices in the test set) and the interpolated data were compared to the ground truth original scans (without missing detectors). 
Between-sinogram and between-image comparisons were based on the SSIM and MAE in Equations (\ref{eq:SSIM})-(\ref{eq:MAE}). 
Note that high values are better for the SSIM, while low values of the MAE are the best. 

To complement these global metrics, we measured mean activity ($\mu_{ROI}$) and standard deviation ($\sigma_{ROI}$) in a 5-10 cm$^3$ bladder region of interest (ROI) and a randomly placed background ROI of similar size.
We define background variability (BV) for a ROI as: 
\begin{align*}
BV = \frac{\sigma_{ROI}}{\mu_{ROI}}\cdot 100 \,\%. 
\end{align*}. 
In this study, we report a relative background variability (rBV), defined as the ratio of BV in the restored images to BV in the corresponding original images. 
We further evaluated the contrast recovery (CR) defined as the ratio between the contrast in the restored and the original images, defined as:
\begin{align*}
CR = \frac{\mu_{\text{bladder,restored}} - \mu_{\text{background,restored}}}{\mu_{\text{bladder,original}} - \mu_{\text{background,original}}} \cdot 100 \,\%.
\end{align*}

All data processing steps, including list mode alterations, binning into sinograms, and image reconstructions, were performed with GE Healthcare's PET research toolbox Duetto (v02.18). 
Following the clinical protocol, we performed image reconstruction to 192×192×89 voxels, 60 cm radial and 25 cm axial FOV via the 3D ordered subset expectation maximisation algorithm with 28 subsets, 2 iterations, and a 5mm Gaussian postfilter. The reconstruction used the default normalisation file and corrections for attenuation, scatter and randoms.

For a robust and quantitative comparison, two statistical tests were chosen based on the nature of the data and the aspects of the prediction methods being evaluated. 
Fisher's Z-transformation was used to compare pixel-wise correlations between original and predicted sinograms. 
The Mann-Whitney U test was employed to evaluate significant difference between evaluated measures in both sinograms and reconstructed images. 

\section{Results}\label{sec3}
This section will refer to the ground truth original scans as "original" and the sparse PET configuration-modified data as "distorted". 
Predictions resulting from the sinogram restoration network are referred to as the "restored" sinograms, as opposed to the "interpolated" sinograms that served as a baseline method to fill in the missing data. 
\begin{figure*}[h!]
    \centering
    \captionsetup{width=.91\linewidth}
    \includegraphics[width=1\linewidth]{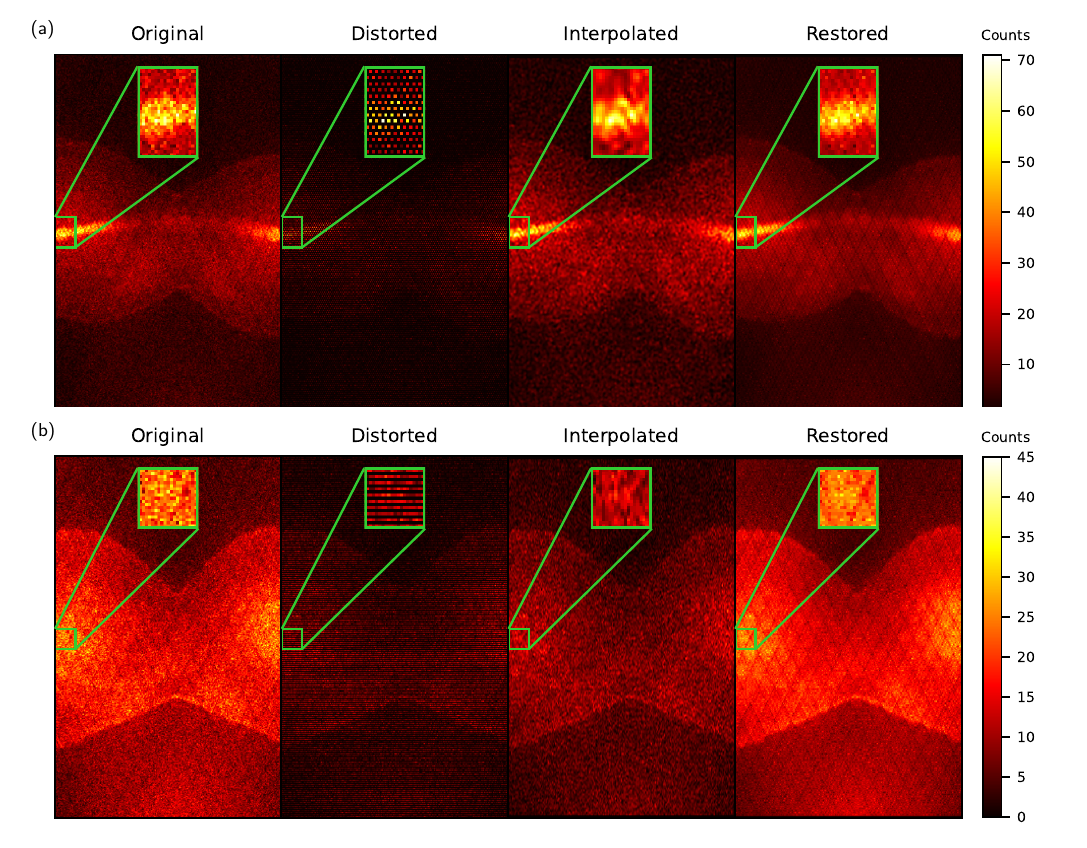}
    \caption{Visual comparison of 2D sinograms based on the ground truth (Original),
        the sparse chessboard configuration (Distorted), 
        and predictions from interpolation filling (Interpolated) 
        and from the sinogram restoration network (Restored). 
        The figure includes two different sinogram slices from one scan in the test set.
         \textbf{(a)} shows a direct plane sinogram slice. 
         \textbf{(b)} shows a summed cross-plane sinogram slice with ring difference 1. 
         \label{fig:Sinos2}
        }
\end{figure*}

\subsection{Sinogram restoration}
\begin{figure}[h!]
    \centering
    \captionsetup{width=.91\linewidth}
    \includegraphics[width=.55\linewidth]{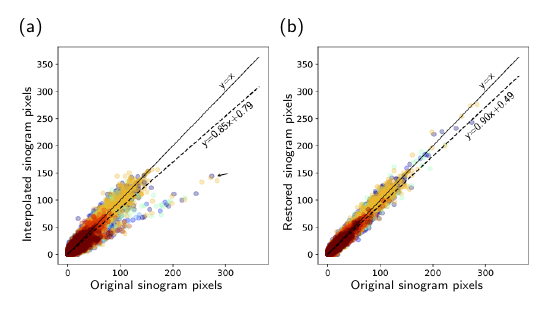}
    \caption{Pixel-wise correlation between the original sinograms and the predicted sinograms from \textbf{(a)} interpolation
        and \textbf{(b)} the restoration network. 
        The figure shows a random sample of $10^5$ pixels from the different scans (different colours for different scans). 
        The ideal fit is illustrated by the solid lines, and the overall total fit to the original pixels is illustrated by the dashed lines.  
        The interpolated pixels exhibit traces of the two separate distributions of the distorted pixels with remaining pixels of lower intensity than the original (marked by arrow).
        \label{fig:Sino_scatterplots_out}
        }
\end{figure}

\begin{figure}[h!]
    \centering
    \captionsetup{width=.91\linewidth}
    \includegraphics[width=.55\linewidth]{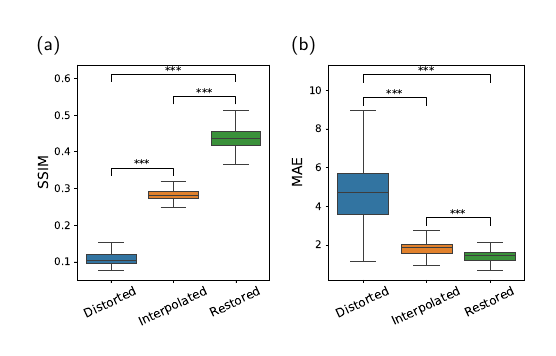}
    \caption{\textbf{(a)} Structural similarity index measure (SSIM) and 
        \textbf{(b)} mean absolute error (MAE) of sinograms from the sparse chessboard configuration (Distorted), 
        and predictions from interpolation filling (Interpolation) and from the sinogram restoration network (Restored). Note that high values are better for the SSIM, while low values of the MAE are the best. 
        Whiskers, box limits, and inner line indicate 5th and 95th percentiles, 25th and 75th percentiles, and median respectively.
        The figure includes results from Mann-Whitney U test for difference between the samples, where *** indicates a $p$-value below 0.001.
        \label{fig:Sino_boxplots}
        }
\end{figure}

The proposed sinogram restoration model successfully recovers missing counts in the distorted sinograms (Figure~\ref{fig:Sinos2}). 
The sinogram restoration network fills in missing counts with high fidelity and demonstrates a strong positive linear correlation with the original counts (Figure~\ref{fig:Sino_scatterplots_out}(b)). 
The restored sinograms cannot reproduce finer details that would have occurred in missing detectors, which inevitably leads to a smoothing effect (Figure~\ref{fig:Sinos2}(b)). 
This is reflected in moderately high SSIM values, as seen in  Figure~\ref{fig:Sino_boxplots}(a). 
The mean absolute pixel deviation is consistently within two counts per pixel (Figure~\ref{fig:Sino_boxplots}(b)).


The interpolated sinograms also present low MAE (Figure~\ref{fig:Sino_boxplots}(b)) and a strong positive correlation with the original counts (Figure~\ref{fig:Sino_scatterplots_out}(a)). 
However, the correlation is significantly lower compared to the restored predictions (${p<0.001}$). 
The interpolation method fails to recover the lower counts in the diagonal cross planes of the distorted data, leading to distinctive underrepresentation in the added cross-plane sinograms for ring differences equal to one (Figure~\ref{fig:Sinos2}(b)). 
The shortcoming of the interpolation method is evident, with inferior values of the SSIM and MAE (${p<0.001}$), as seen in Figure~\ref{fig:Sino_boxplots}. 

Figure~\ref{fig:Sinos2} shows representative 2D direct sinograms and summed cross-planes with ring difference 1. The interested reader is referred to supplementary Figure~\ref{fig:supp_sinos} for corresponding examples from the same scans for larger ring differences (5 and 11), confirming similar cancellation patterns and restoration behaviour across oblique planes.

\subsection{Reconstructed images}
\begin{figure*}[h!]
    \centering
    \captionsetup{width=.91\linewidth}
    \includegraphics[width=1\linewidth]{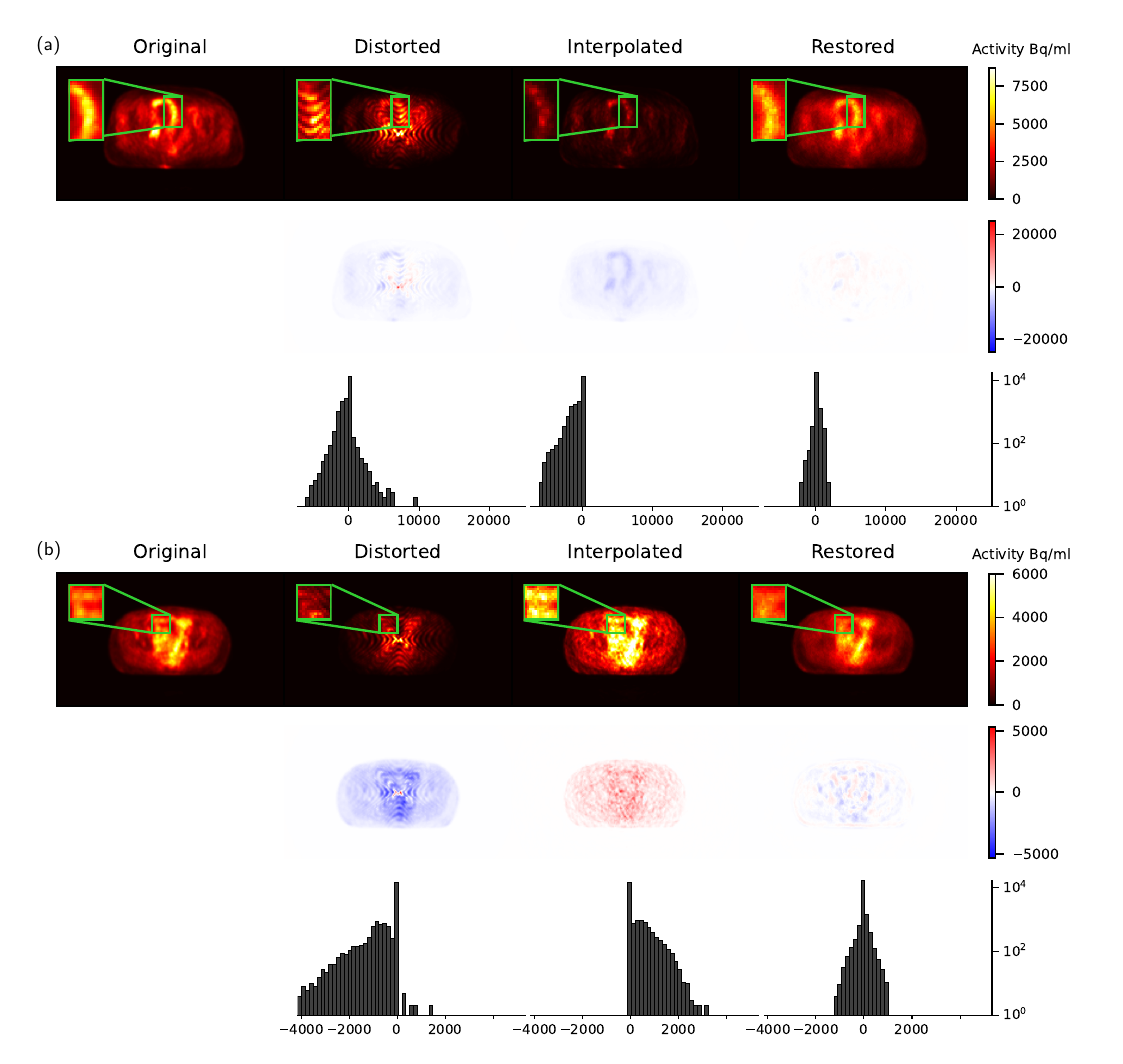}
    \caption{Visual comparison of reconstructed image slices based on original sinograms (Original), sinograms from the sparse chessboard configuration (Distorted), 
        and predicted sinograms from interpolation (Interpolated) and from the restoration network (Restored). 
        \textbf{(a)} and \textbf{(b)} show two different image slices from two different scans in the test set, with reconstructed images on the top vertical panel, corresponding difference images on the middle panel (colour bar in Bq/ml), and corresponding histograms of pixel differences in the bottom panel ($x$-axis in Bq/ml). 
        \label{fig:Recons2}
        }
\end{figure*}

\begin{figure}[h!]
    \centering
    \captionsetup{width=.91\linewidth}
    \includegraphics[width=.55\linewidth]{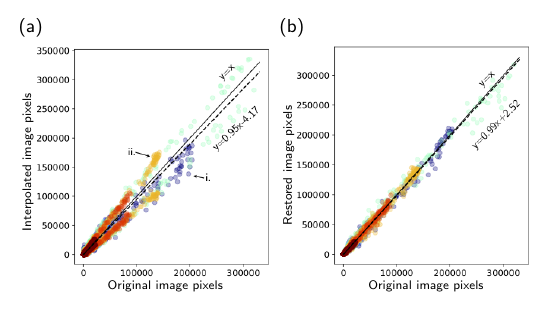}
    \caption{Pixel-wise correlation between the original images and images reconstructed based on the predicted sinograms from the restoration network 
        from \textbf{(a)} interpolation
        and \textbf{(b)} the restoration network.
        The figure shows a random sample of $10^5$ pixels from the different scans (different colours for different scans). 
        The ideal fit is illustrated by the solid lines, and the overall total fit to the original pixels is illustrated by the dashed lines. 
        The interpolated image pixels indicate two separate distributions of the image pixels with either lower (i) or higher (ii) intensity than the original (marked by arrows).
        \label{fig:Recon_scatterplots_out}
        }
\end{figure}

How the effects of sinogram restoration translate in terms of quality of reconstructed images is shown in Figure~\ref{fig:Recons2}. The smooth nature of the restored sinograms leads to reconstructed images with smoother background texture and some lack of finer details. This is manifested as somewhat reduced high-contrast regions and missing low-contrast regions (Figure~\ref{fig:Recons2}). Quantitatively, this was reflected in the background variability, where low-contrast regions appeared slightly noisier than the original images (rBV$\approx$1.08), while high-contrast regions without specific bindings such as the bladder were slightly smoother (rBV$\approx$0.92). A similar pattern can be seen in the pixel-wise deviation (Figure~\ref{fig:Recon_scatterplots_out}(b)). The main deviation is attributed to the high-contrast pixels and has an underestimating effect.

\begin{figure}[h!]
    \centering
    \captionsetup{width=.91\linewidth}
    \includegraphics[width=.55\linewidth]{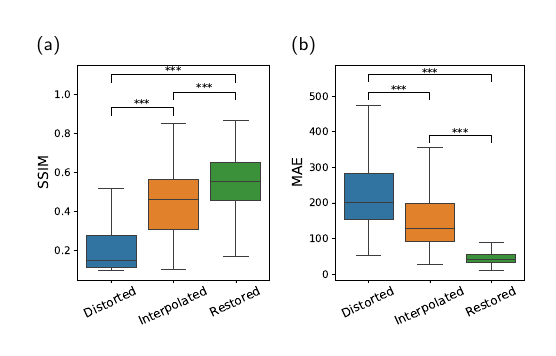}
    \caption{\textbf{(a)} Structural similarity index measure (SSIM) and 
        \textbf{(b)} mean absolute error (MAE) of reconstructed images based on the sinograms from the sparse chessboard configuration (Distorted), 
        and predictions from interpolation filling (Interpolation) and from the sinogram restoration network (Restored). Note that high values are better for the SSIM, while low values of the MAE are the best. 
        Whiskers, box limits, and inner line indicate 5th and 95th percentiles, 25th and 75th percentiles, and median respectively.
        Figure includes results from Mann-Whitney U test for difference between the samples, where *** indicates a $p$-value below 0.001.
        \label{fig:Recon_boxplots}
        }
\end{figure}

The underestimation in interpolated sinogram counts seen in  Figure~\ref{fig:Sinos2}(b) has a pronounced impact on reconstruction, resulting in image slices with lower intensity (Figures~\ref{fig:Recons2}(a) and \ref{fig:Recon_scatterplots_out}(a)). 
Moreover, the interpolated sinograms contribute to occasional spurious features in the reconstructed images, appearing as overestimated intensities, visible in  Figures~\ref{fig:Recons2}(b) and \ref{fig:Recon_scatterplots_out}(a). 
This tendency is reflected in the histograms in Figure~\ref{fig:Recons2}, where the interpolated images exhibit a broader and more uneven distribution of pixel differences relative to the original images. 
In contrast, the restored images show a narrower error distribution, indicating closer overall agreement with the original, despite the inherent smoothing.
The visual comparison is also appreciated in the quantitative measures in  Figure~\ref{fig:Recon_boxplots}, with superior values of SSIM and MAE for the restored images (${p<0.001}$). 

Image slices from the same scans of Figure~\ref{fig:Recons2} but reconstructed in coronal view can be found in supplementary Figure~\ref{fig:supp_recons}, supporting the same qualitative trends while revealing some band-shaped artefacts, most evident in the interpolated images.


\section{Discussion}\label{sec4}
This study presents a novel method for deep learning-based filling of incomplete sinograms resulting from sparse PET configuration systems with inter-detector gaps following a chessboard pattern. 
Motivated by other studies on sparse detector configurations~\citep{Daube-Witherspoon2021, Karakatsanis2022, Knopp2019, ZeinKarakatsanis2020, Yamaya2008, Yamaya2009, AbiAklVandenberghe2019, ZeinKarakatsanis2021, Karakatsanis2018, Karakatsanis2019, Zhang2019}, we explored several patterns for detector cancellation. 
In addition to the expected sensitivity loss from missing counts relative to non-sparse designs~\citep{AbiAklVandenberghe2019, ZeinKarakatsanis2021}, specific image quality degradations have been reported with increasing gap sizes. 
Background variability and losses in spatial resolution and contrast recovery tend to increase with the axial gap size~\citep{Daube-Witherspoon2021, Karakatsanis2022}. 
Other known drawbacks are noise artefacts due to axial gaps and streaking artefacts due to transaxial gaps~\citep{Yamaya2008, Yamaya2009, Sheikhzadeh2019}. 
The chessboard pattern used here was chosen to generate maximum detector sparsity (50\% of the detectors) with minimal gap size. 
We investigated the data loss effects of several block sizes for the chess squares. 
A preliminary evaluation of the sinogram restoration network on configurations of sizes $1\times1$, $2\times2$, and $3\times4$ showed robustness in terms of SSIM and MAE (data not shown). 
However, only the best-performing (in terms of SSIM and MAE) chessboard pattern size $1\times1$ was fine-tuned and included in the presented results. 

The restoration network successfully recovered distorted sinograms. However, the smooth nature of the restored sinograms causes excess smoothing also in the reconstructed images, limiting clinical usability at this point. Although small high-contrast regions are detected in the images, the model demonstrates a lack of precision in correctly quantifying these areas. The smoothing effect seems to have a direct impact on the contrast recovery, which was evident in the reconstructed image in Figure~\ref{fig:Recons2} where some high-contrast areas are reduced in size and some are distorted in their shape. It is also evident in the pixel-wise contrast comparison in Figure~\ref{fig:Recon_scatterplots_out}(b), where we see an indication of high-contrast pixels as the main drivers to the deviation from a perfect fit to the original pixels.

In contrast, the interpolation method performed worse, underestimating specific sinogram slices and overestimating some pixels, causing hallucinatory artefacts in some reconstructed image slices. 
This is further evidenced by the histograms in Figure~\ref{fig:Recons2}, which show that the interpolated method results in wider and more irregular error distributions, reinforcing its tendency toward inconsistent pixel-level accuracy.
The underestimated sinogram slices coincide with certain cross planes, adding up oblique angles from two directions and a position shift in the distortion pattern, resulting in lower counts across all pixels. Interpolation cannot infer systematic count loss but only fills in gaps, often leading to artefacts.

The 2D sinograms corresponding to summed cross planes with ring difference 1 displayed both zeroed-out lines and overall lower counts values in the remaining pixels and, therefore, differ from the ground truth in all sinogram pixels (Figure~\ref{fig:Sino_scatterplots_in}(b)). Hence, the sinogram restoration network had to learn to predict entire sinogram slices for these cross planes. This approach is not possible for the interpolation, where reference points are needed to generate predictions. Therefore, a different mask was supplied to the interpolation method for these mentioned cross planes, leaving the low-count pixels as reference points, resulting in an overall underestimation of the zeroed-out pixels. Even when boosted with a global scaling factor, the interpolation approach consistently underestimated counts and could not match the performance of our restoration network. 
The effect is likely responsible for the band-shaped artifacts observed in the reconstructed coronal images (Figure~S.2). 
This behaviour may be related to the 2D formulation adopted in this work, where the network was trained on individual sinogram slices without explicit modelling of inter-slice correlations. 
Further investigation into architectures that better capture cross-plane dependencies may therefore be warranted.

Building on these observations, future work should explore extending the current 2D network into a 3D formulation to explicitly model volumetric context, as well as an investigation of incorporating sharpness to the U-Net structure to mitigate the smoothing effect inherent in neural networks. It would also be interesting to investigate this specific model's capacity regarding possible data losses for clinically approved image quality and if there is an optimal sampling pattern. One might consider an adaptive sparsity pattern, e.g., tighter sampling at the scanner centre and more sparse sampling at the scanner ends. Another necessary step is to explore the integration of sinogram re-normalisation to account for the missing lines of response, particularly in the cross-planes. In this study, we did not explicitly apply re-normalisation. However, our results show that the network is capable of restoring both the missing signals in the direct planes and the attenuated signals in the cross-planes without a separate re-normalisation. 
This suggests that the model can, to some extent, compensate for the effects of missing and attenuated signals without explicit need for re-normalisation of the sinograms. 
Nonetheless, applying explicit re-normalisation as a preprocessing step may reduce the network's burden and improve restoration accuracy -- a possibility that warrants further investigation. 

Furthermore, the dataset was limited in the sense that it only included one anatomical region and only left one patient for validation. This was a main motivation for training the model on individual sinogram slices, which comes at a possible trade-off of introduced bias in the results. The current model was furthermore trained on one specific detector pattern at a time. To increase model generalisability, the dataset should be extended to multiple anatomical regions, and the training process could be extended to include several pattern sizes. 
Our model could also be combined with parallax correction. In the case of missing LORs due to the missing detector elements, such a correction scheme would also give the added benefit of shifting some of the measured counts to empty detector positions, thereby assisting the estimation of missing counts. 
Another alternative is to extend our model to include time-of-flight (TOF) information. We expect that including the TOF timing resolution would improve lesion detectability compared to the non-TOF setting due to enhanced signal-to-noise, mainly due to the increase in the number of 2D sinograms available for training. We do not expect the reported blurring to go away since the same detector elements are still missing. 
An additional future direction would be to evaluate whether combining our method with, for instance, bed motion could enhance sinogram restoration, since bed motion effectively provides multiple slightly shifted sampling configurations of the same patient position. 
In summary, the present work should be regarded as a proof-of-concept from a deep learning standpoint, with practical implementation requiring further development.

The proposed sparse chessboard configuration is designed to explore the potential for lower-cost or more cost-effective clinical PET scanners, aiming for the same axial FOV as conventional scanners but with only 50\% of the original detector materials. 
We further speculate that the proposed sparse geometry could alternatively be used to allow the development of a two-fold extension of the axial FOV of conventional compact ring configurations with the same number of detectors. An extended axial FOV would, for instance, allow simultaneous imaging of distant organs such as the brain and heart, as in the total body PET systems available on the market. 
This enhanced coverage enables static whole-body imaging, a challenge with standard systems due to axial limitations.
A sparse long axial FOV PET scanner with 50\% fewer detectors, retaining only 25\% of the initial LORs, would inevitably have substantially lower sensitivity than full-coverage total-body scanners. However, its sensitivity would remain comparable to that of earlier-generation PET systems, which have more limited axial coverage. 
From this perspective, such a system could still enable whole-body imaging at a reasonable cost and acceptable sensitivity. 
Combining the two approaches of extended axial FOV and reduced detector cost compared to compact designs results in a medium-cost extended FOV, which offers many of the advantages of a full total body PET but positioning itself between the current conventional PET and extended FOV systems in terms of investments. We note, however, that while the 1×1 sparsity pattern performed best in our evaluation, such a configuration may not be the most practical or cost-effective from a manufacturing perspective.

Finally, although our work focuses on sparse ring-based PET configurations, other approaches share the same goal of reducing detector costs through novel system geometries. Flat-panel PET configurations~\citep{vandenberghe2023flatpanel, abi2025flatpanel}  reflect the same broader objective of enabling affordable PET imaging through hardware sparsity. Preliminary studies have applied deep learning-based denoising to flat-panel PET systems with sparse detectors, showing promising image quality with a 30\% detector reduction~\citep{abiakl2024deep-learning}.

\section{Conclusion}\label{sec5}
This study serves as a proof of concept that sparse PET configurations with inter-detector gaps are within reach with the aid of deep learning-based data restoration. 
Our model demonstrated promising results in sinogram restoration.
However, certain limitations, particularly in preserving finer details, indicate the need for further refinements. 
Nevertheless, this research represents a step forward in increasing the availability of clinical PET systems worldwide by realising the practical potential of deep learning approaches for designing low-cost extended FOV PET systems.

\section*{}
\vspace*{-10mm}
\subsection*{Ethics approval and consent to participate}
The studies involving humans were approved by The Regional Ethical Review Board, Umeå (dnr 2015/117-31). The studies were conducted in accordance with the local legislation and institutional requirements. The participants provided their written informed consent to participate in this study.

\vspace*{-5mm}
\subsection*{Data availability statement}
The dataset and code used for this study can be provided by the corresponding author upon reasonable request.

\vspace*{-5mm}
\subsection*{Acknowledgments}
The authors acknowledge the use of high-performance computing resources provided by High Performance Computing Center North (HPC2N), which supported the computational work for this study. 
This work is supported by the Swedish Research Council (340-2013-5342).

\vspace*{-5mm}
\subsection*{Conflict of Interest Statement}
The authors declare that the research was conducted in the absence of any commercial or financial relationships that could be construed as a potential conflict of interest.

\bibliographystyle{unsrtnat}
\bibliography{main.bib}

\newpage
\section*{Supplementary Materials}

\vspace*{2cm}
\setcounter{figure}{0}
\renewcommand{\thefigure}{S.\arabic{figure}}

\begin{figure*}[h!]
    \centering
    \captionsetup{width=.91\linewidth}
    \includegraphics[width=0.92\linewidth, trim={0mm 5mm 0mm 5mm}]{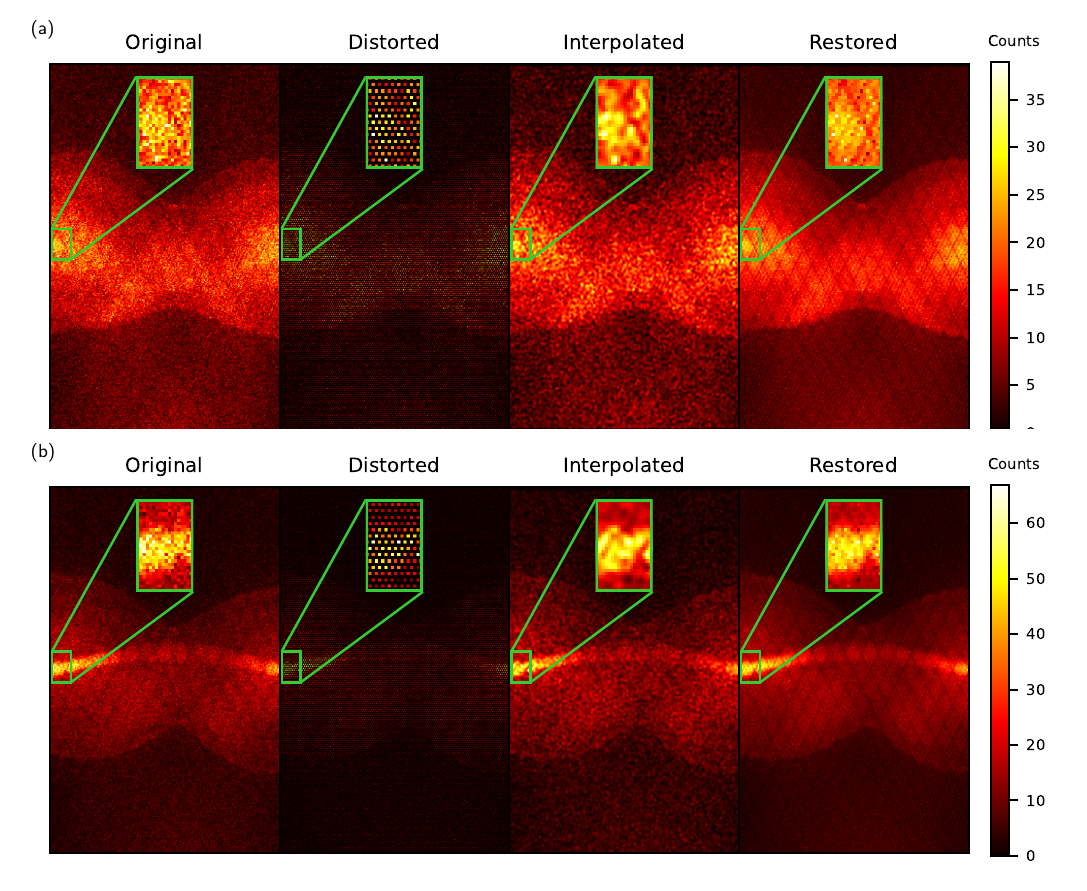}
    \caption{Visual comparison of 2D sinograms based on the ground truth (Original), the sparse chessboard configuration (Distorted), and predictions from interpolation filling (Interpolated) and from the sinogram restoration network (Restored). The figure includes two different sinogram slices from one scan in the test set. 
    \textbf{(a)}  shows a sinogram slice with ring difference 5, and 
    \textbf{(b)} shows a sinogram slice with ring difference 11.
    \label{fig:supp_sinos}
    }
\end{figure*}

\begin{figure*}[h!]
    \centering
    \captionsetup{width=.91\linewidth}
    \includegraphics[width=1\linewidth, trim={0mm 0mm 0mm 0mm}]{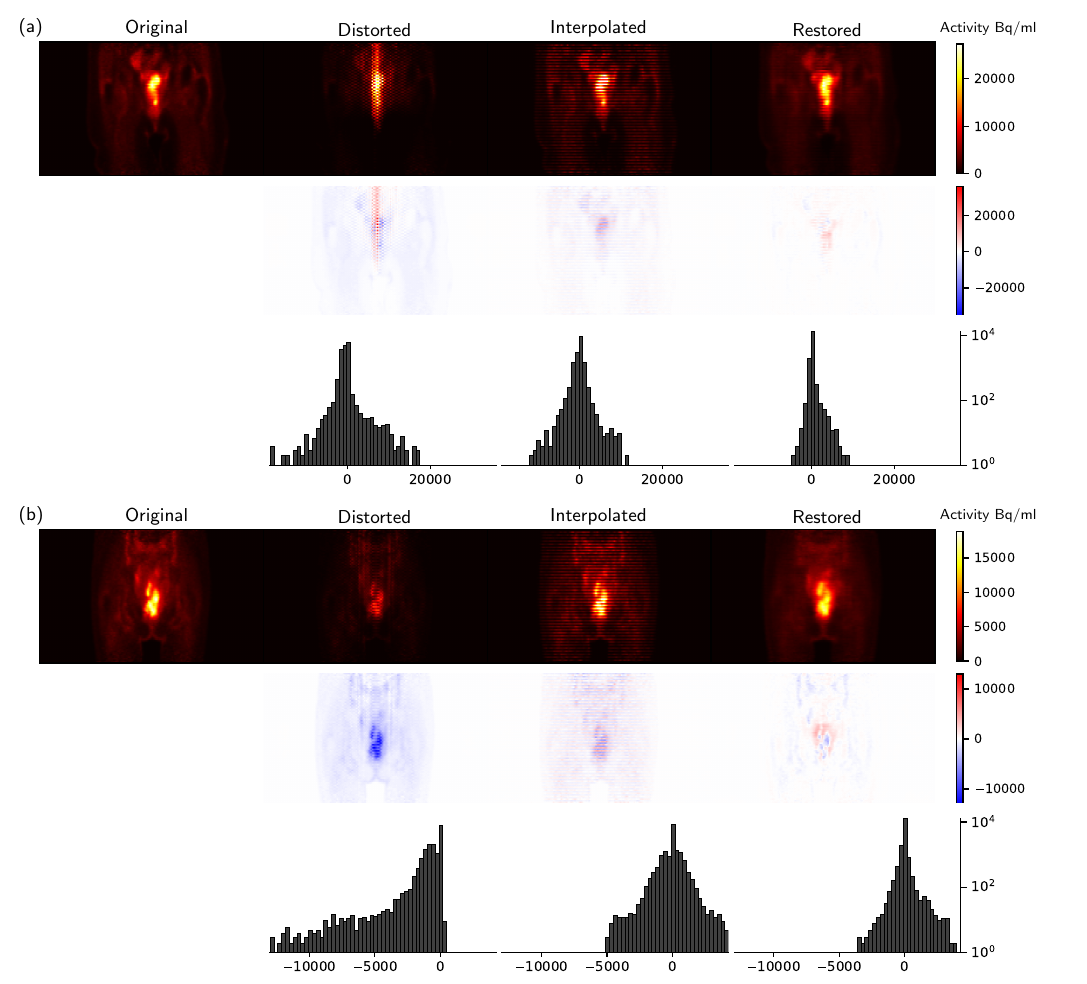}
    \caption{Visual comparison of reconstructed image slices based on original sinograms (Original), sinograms from the sparse chessboard configuration (Distorted), and predicted sinograms from interpolation (Interpolated) and from the restoration network (Restored). 
    \textbf{(a)} and \textbf{(b)} show two different image slices from two different scans in the test set, with reconstructed images in the coronal view on the top vertical panel, corresponding difference images on the middle panel (colour bar in Bq/ml), and corresponding histograms of pixel differences in the bottom panel ($x$-axis in Bq/ml).
    \label{fig:supp_recons}
    }
\end{figure*}

\end{document}